# Connected C-Core Hybrid SRMs for EV Applications

Gholamreza Davarpanah
Department of Electrical Engineering
Amirkabir University of Technology
Tehran, Iran
ghr.davarpanah@aut.ac.ir

Sajjad Mohammadi
*Department of Electrical Engineering and Computer Science*
*MIT*
Cambridge, USA
sajjadm@mit.edu

*Abstract—* This paper proposes a new class of permanent magnet-assisted three-phase switched reluctance motors (PM-SRMs) designed to achieve significantly higher torque density for electric vehicle (EV) propulsion systems. Eight distinct motor topologies are systematically investigated, including a non-PM baseline design, three innovative PM arrangement strategies, and two optimized rotor/stator teeth configurations (22-pole and 26-pole variants). The study presents analytical models including magnetic equivalent circuits (MECs), detailed operating principles, and generalized design formulations that account for both electromagnetic and structural considerations. A key contribution is the introduction of the point-of-conversion (PoC) concept, which optimizes PM placement by minimizing magnetic path reluctance. Comparative analysis demonstrates torque density improvements over conventional SRMs and existing PM-assisted designs while maintaining structural robustness. Experimental validation confirms that the proposed 24/22 configuration with inter-phase PMs delivers higher torque per PM volume compared to state-of-the-art designs. The findings provide insights for EV motor designers seeking to balance performance, cost, and reliability.

## I. Introduction

Switched reluctance motors (SRMs) have emerged as a promising solution for electric vehicle powertrains due to their inherent robustness and fault-tolerant characteristics [1]-[5]. Recent advancements in SRM design have focused on torque density enhancement [6]-[11], with various topological approaches being explored. Parallel developments in E-core SRM designs have shown promising torque characteristics [12], particularly in their connected-core variants, which offer enhanced mechanical strength [13], [14]. [15] Introduces an innovative design equation that systematically determines the optimal stator-to-rotor teeth ratio, enabling high-performance SRM configurations with strategically increased rotor teeth for enhanced torque characteristics. To achieve higher output torque and improved efficiency in switched reluctance motors (SRMs), the incorporation of permanent magnets (PMs) as auxiliary excitation has emerged as an effective solution. This hybrid approach combines the inherent advantages of SRMs with enhanced torque characteristics, making them increasingly viable alternatives to conventional PM motors. These merits have positioned PM-assisted SRMs as promising candidates for traction motor applications in electric vehicles and other transportation electrification systems [16]-[21]. While C-core configurations with PMs between C-cores of different phases demonstrate torque improvement [22], the back PMs are not very effective, and the unwanted radial forces negatively impact long-term operation of the motor. [23] introduces a modular SRM with an A-type stator, incorporating PMs within the stator back iron. In [24], an outer-rotor 24/26 HESRM with PMs embedded between C-cores of different phases is comprehensively analyzed, yet it could face structural issues as we have bearings on one side of the shaft; also, magnet utilization is not effective with the designed stator/rotor teeth combination. This paper presents a comprehensive study of three-phase PM-assisted SRMs, introducing eight novel topologies with optimized magnet placement strategies and stator/rotor teeth combinations. The work includes detailed operating principles analysis, development of magnetic equivalent circuits, and thorough experimental validation of prototyped designs, significantly advancing high-torque-density SRM technology.

## II. Topologies and Design Techniques

Fig. 1 illustrates the eight proposed motor topologies, systematically designed with two distinct rotor teeth configurations (22 and 26 poles) and four permanent magnet arrangements: the baseline non-PM configuration (1a and 1b), PM1 placement between C-cores of different phases (2a and 2b), PM2 (inside the C-core of each phase (3a and 3b), and the combined PM1+PM2 (4a and 4b). These design variations present important engineering trade-offs - while increasing rotor teeth count enhances torque potential, it simultaneously reduces pole width, resulting in three key challenges: elevated core losses due to increased magnetic flux density, decreased commutation angles affecting control precision, and reduced available winding slot area limiting electrical capacity. As demonstrated in reference [20], comprehensive optimization reveals an optimal number of teeth of two for balancing these competing factors. The flux distribution analysis shows that the majority of PM-generated flux effectively crosses the airgap to participate in torque production. At the same time, a small portion circulates through the C-core structure in opposition to coil-generated flux. This flux opposition phenomenon creates an opportunity to increase winding area while maintaining equivalent magnetic loading, thereby enabling higher electrical loading capacity. To improve structural robustness, c-cores of the same phase are diagonally placed on opposite sides of the rotor to cancel out radial forces. To enhance mechanical robustness, c-cores of the same phase are diagonally placed on opposite sides of the rotor. This symmetrical arrangement effectively cancels out unbalanced radial forces.

## III. Operating Principles and Magnetic Circuits

Fig. 2 illustrates the flux behavior characteristic of the proposed PM-assisted SRM. Under zero-current conditions, the permanent magnet flux primarily circulates through the stator yoke, completing its magnetic circuit without significant airgap penetration, thereby making a negligible contribution to torque generation. However, as stator current increases, three important phenomena occur simultaneously: (1) the C-core material begins to approach magnetic saturation, (2) the resulting permeability reduction forces PM flux to redirect across the airgap, and (3) this redirected flux actively participates in the energy conversion process, producing measurable torque enhancement. The higher the current, the greater the contribution of the PM to provide the torque. The MEC of the motors is derived as shown in Fig. 3. The most effective strategy is to place the PMs as close to the air-gap where the energy conversion is taking place. It can be called point-of-conversion (PoC) PM placement, which suggests placing the magnets close to the air-gap. This strategic placement provides two key advantages: (1) it minimizes the magnetomotive force (MMF) drops along iron paths; (2) it ensures maximum flux linkage with the stator windings. The PoC methodology represents a significant performance improvement over conventional PM positioning approaches, as it specifically targets the regions of highest magnetic energy conversion efficiency.

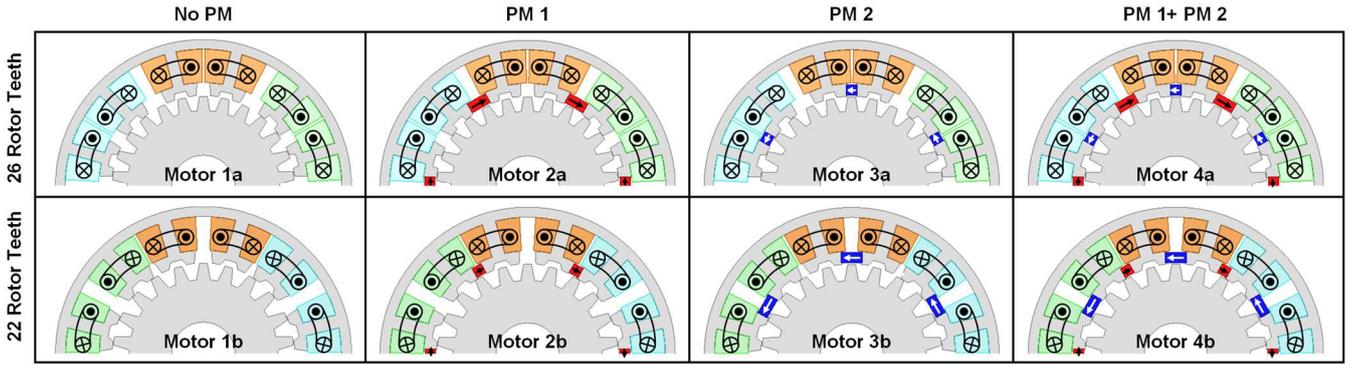

Fig. 1. Topologies of the eight SRMs.

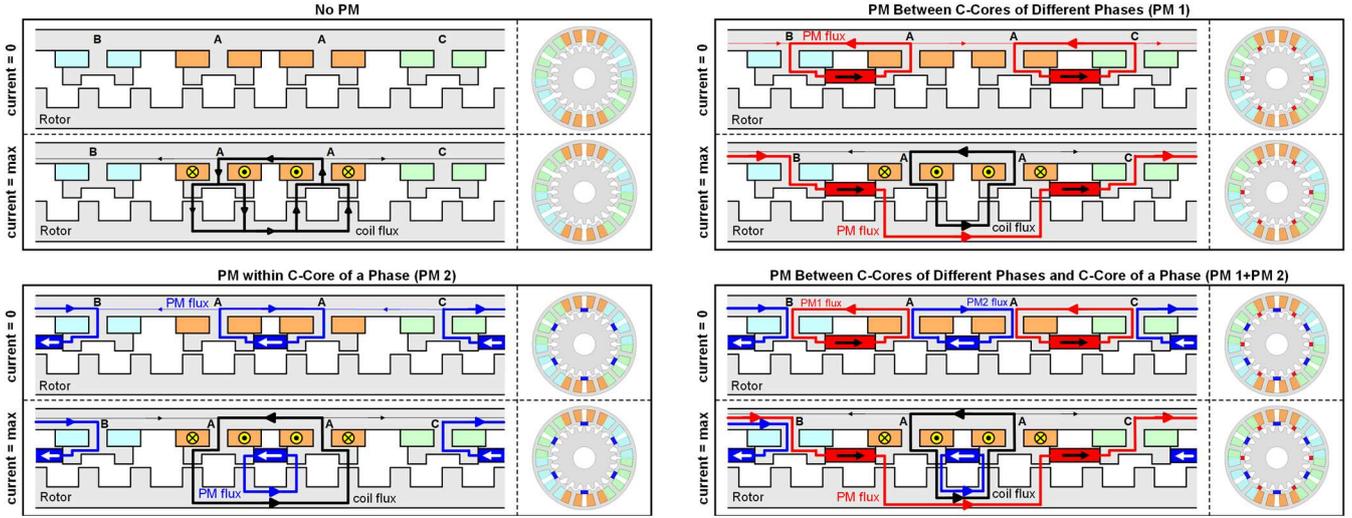

Fig. 2. Magnetic flux paths of the studied SRMs: (a) No PM, (b) PM1, (c) PM2, and (d) both PM1 and PM2.

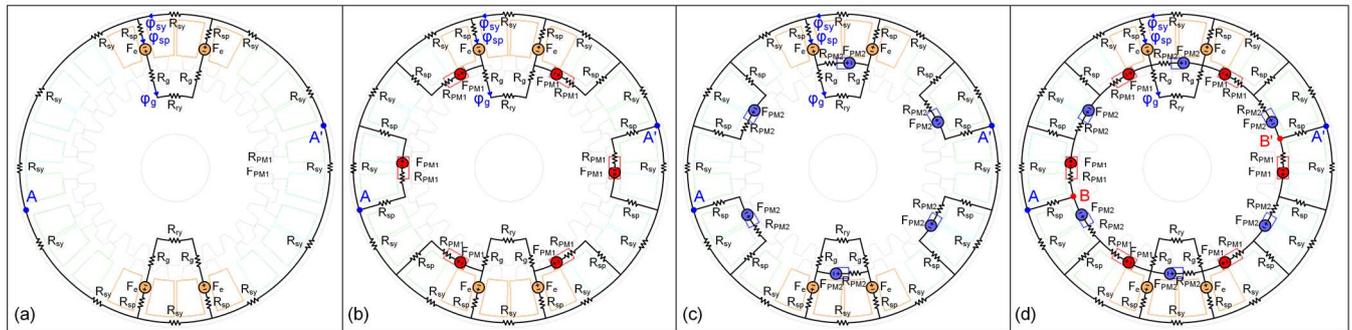

Fig. 3. Magnetic equivalent circuits of motors (a) 1a and 1b, (b) 2a and 2b, (c) 3a and 3b, and (d) 4a and 4b.

IV. SIMULATION RESULTS AND DISCUSSIONS

Fig. 4 shows the flux distribution in all eight motor designs at aligned and unaligned positions under 8 A excitation. The PM-integrated designs exhibit reduced saturation in stator poles and yoke compared to non-PM versions, demonstrating more efficient flux control. Fig. 5 illustrates the total torque output and the individual contributions from the coils and the PMs. The results clearly show that while PM placement enhances overall torque production, simply increasing magnet volume does not always yield higher torque. A prime example is motor 2b (24/22 with PM1 arrangement), which—despite using 50% less PM material—achieves greater torque output

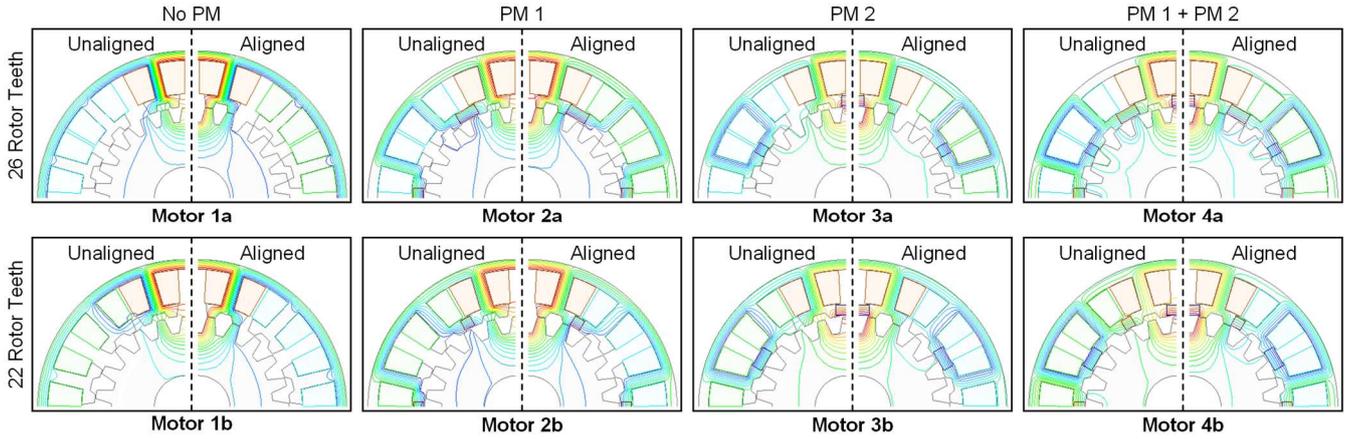
Fig. 4. Flux lines within SRMs in aligned and unaligned conditions.

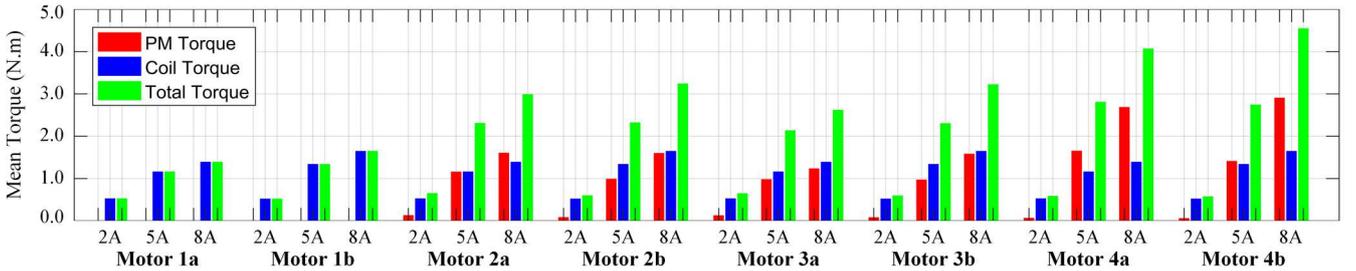
Fig. 5. Decomposition of total torque components to contribution of the stator coil and the PMs.

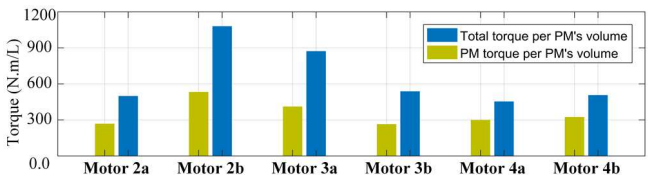
Fig. 6. Torque per magnet volume.

than motor 2a (24/26 with PM1 set, comparable to [23]). This efficiency is further quantified by the torque-per-PM-volume index depicted in Fig. 6. Additionally, motor 3a (24/26 with PM2 configuration) outperforms motor 3b (24/22 with PM2 set) in torque generation while demonstrating superior PM utilization efficiency, reinforcing that strategic magnet placement outweighs sheer PM quantity. To ensure fair comparison, all motor designs—including both conventional and hybrid configurations—were redesigned to maintain identical physical dimensions, air gap specifications, and material properties. Additionally, all motors were optimized using the same methodology applied to the proposed motor. The same optimization methodology was applied to all motors and our proposed motor design. This stringent standardization process guarantees that all performance comparisons—including mean torque, torque density, torque-per-ampere, and torque per PM volume—are conducted on the same level.

Our comprehensive evaluation compares the proposed hybrid excitation SRM (Motor 4b) against two baseline SRM designs, three existing HESRM from prior work [17], [19], [21], and a competitive flux-switching permanent magnet (FSPM) motor from reference [25]. The performance data presented in Table I leaves no doubt regarding the superior capabilities of our proposed motor, which consistently outperforms all comparator motors across every performance metric.

## V. EXPERIMENTAL RESULTS AND DISCUSSIONS

Fig. 7 illustrates the prototypes of both 24/22 and 24/26 SRM configurations, showing the integration of permanent magnets between the C-core teeth. This design not only enhances electromagnetic performance but also significantly improves the structural integrity of the stator assembly. Four small appendices fasten these PM modules in position, adding minimal complexity to the manufacturing process while ensuring reliable operation. The experimental configuration shown in Fig. 8 comprises a torque transducer, an optical rotary encoder for position feedback, and a computer-controlled stepper motor system for accurate angular positioning of the test specimens. Fig. 9 demonstrates the measured torque-angle characteristics for all eight motor variants operating at 8A phase current, revealing remarkable consistency with finite-element simulation predictions and thereby validating both the design approach and implementation.

TABLE I
COMPARISON OF THE STATIC RESULTS OF THE PROPOSED MOTOR WITH OTHER MOTORS, [31], [34], [35].

| Parameter | Motor 4b | 12/8 SRM | 8/4 SRM | HESRM [17] | HESRM [19] | HESRM [21] | FSPM [25] |
|---|---|---|---|---|---|---|---|
| Motor volume *(mL)* | 307 | 307 | 307 | 307 | 307 | 307 | 307 |
| PMs volume *(mL)* | 9 | - | - | 20.00 | 16.80 | 21.91 | 33.35 |
| Current *(A)* | 8 | 8 | 8 | 8 | 8 | 8 | 8 |
| Average torque *(N.m)* | 4.540 | 2.975 | 2.213 | 2.612 | 2.472 | 2.235 | 2.899 |
| Torque density *(N.m/L)* | 14.77 | 9.69 | 7.20 | 8.50 | 8.05 | 7.28 | 9.44 |
| Torque per ampere *(N.m/A)* | 0.567 | 0.37 | 0.27 | 0.32 | 0.31 | 0.28 | 0.362 |
| Torque per PMs volume *(N.m/L)* | 504.44 | - | - | 130.60 | 147.14 | 102.00 | 86.92 |

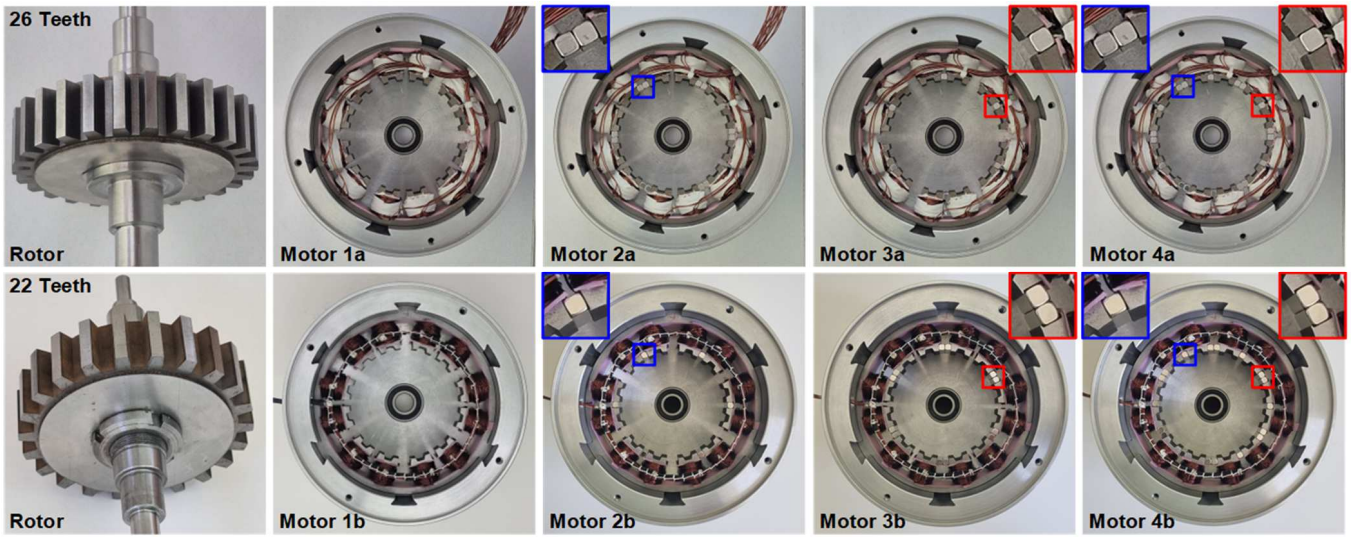

Fig. 7. Prototyped SRMs with $N_r$=26 and $N_r$=22.

## VI. CONCLUSION

This paper presented a new family of C-core hybrid SRMs combining permanent magnet with optimized stator/rotor topologies for electric vehicle applications. The proposed designs demonstrate enhanced torque production through strategic magnet placement and balanced force cancellation, validated by both simulation and experimental results. The point-of-conversion approach proves particularly effective in maximizing PM flux utilization while maintaining structural integrity. Prototype testing confirms the superior performance of the 24/22 configuration with inter-phase PMs, establishing these motors as viable high-performance alternatives for traction systems. The findings provide practical design guidelines for developing efficient, robust, torque-dense SRMs suitable for electrified transportation.

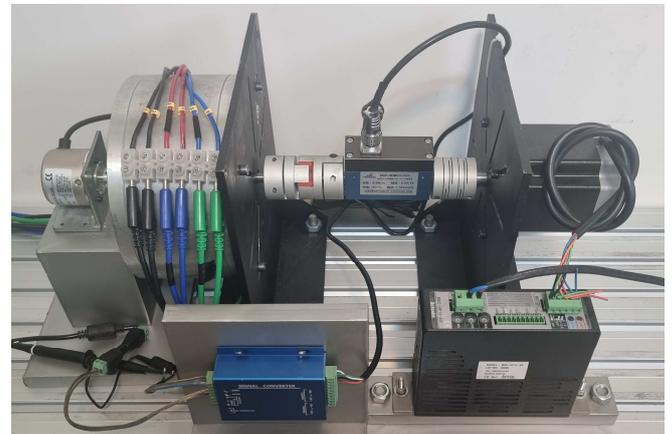

Fig. 8. Torque-angle measurement setup including the SRM, a stepper motor and a torque meter

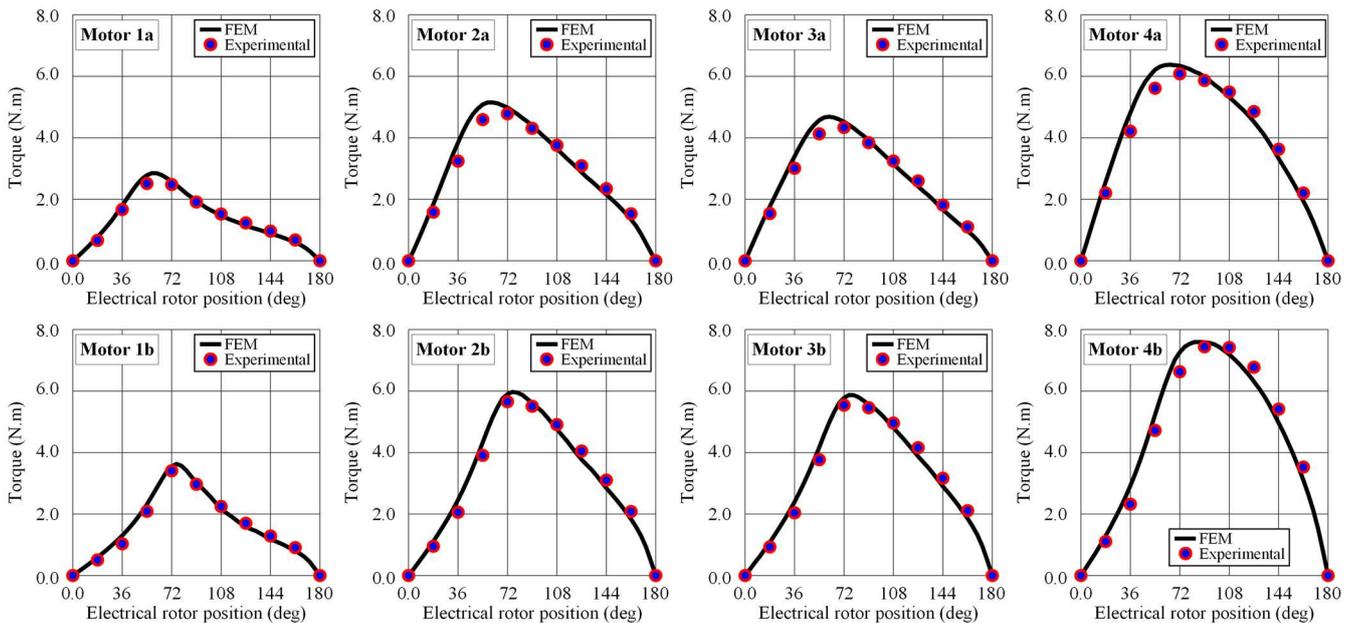

Fig. 9. Comparison of static torque-angle characteristics: simulation versus experimental results.